\documentstyle[amssymb,12pt,thmsa,sw20aip]{article}
\input{tcilatex}
\begin{document}

\title{THE\ CASIMIR\ EFFECT\ UPON\ A\ SINGLE\ PLATE}
\author{Pervez Hoodbhoy \\
%EndAName
Department of Physics\\
Quaid-e-Azam University\\
Islamabad 45320, Pakistan.}
\maketitle

\begin{abstract}
In the presence of an external field, the imposition of specific boundary
conditions can lead to interesting new manifestations of the Casimir effect.
In particular, it is shown here that even a single conducting plate may
experience a non-zero force due to vacuum fluctuations. The origins of this
force lie in the change induced by the external potential in the density of
available quantum states.
\end{abstract}

Externally imposed boundary conditions on a freely fluctuating
electromagnetic field lead to the famous Casimir force between conducting
surfaces separated by some small distance\cite{Casimir}. Recent interest has
been stimulated by improvements in the ability to measure this force, and
many theoretical developments have resulted as well \cite{Milton}. Of
course, it is not necessary to consider just free fields. We could imagine a
situation where the surfaces are embedded inside some classical external
field (such as gravity). Although the interaction of electromagnetism with
gravity is extremely weak, it may nevertheless be interesting to ask how the
force between plates is changed by the external field. As it turns out,
there are some non-trivial consequences. It will be shown in this note that
even a single surface can experience a net non-zero Casimir force under the
influence of a linear external field.

As the simplest possible situation, consider a real scalar field $\phi (x)$
described by the Lagrangian $L=\frac{1}{2}(\partial _{\mu }\phi )^{2}-\frac{1%
}{2}V(x)\phi ^{2}$, where $V(x)$ is an externally prescribed field\footnote{%
Jaffe and co-workers\cite{Jaffe} have used $V(x)$ as a means to mock up the
physical distribution of matter in conducting plates and address questions
relating to conductivity at high frequencies. The purpose of introducing $%
V(x)$ in this paper is different. We note that Elizalde and Romeo\cite
{Elizalde} also considered a one-dimensional system perturbed by an external
field. They did not, however, solve the system under the boundary conditions
used in this paper.}. The Green's function $G(x,x^{^{\prime }})$ obeys $%
(\square +V)=\delta ^{2}(x-x^{^{\prime }})$. In this initial investigation $%
\mu =0,1$ only and $G(x,x^{^{\prime }},k)$, the Fourier-transformed Greens
function, obeys, 
\begin{equation}
\left[ \frac{d^{2}}{dx^{2}}+k^{2}-V(x)\right] G(x,x^{^{\prime }},k)=-\delta
(x-x^{^{\prime }}).  \label{ode}
\end{equation}
The force is readily obtained as the space-space component of the canonical
energy-momentum tensor $T^{\mu \nu }$, 
\begin{equation}
T^{xx}=-i\int_{0}^{\infty }\frac{dk}{2\pi }\left( \frac{\partial }{\partial x%
}\frac{\partial }{\partial x^{^{\prime }}}+k^{2}-V\right) G(x,x^{^{\prime
}},k)\mid _{x=x^{^{\prime }}}.  \label{force}
\end{equation}
If one sets $V=0$ and imposes the Dirichlet condition at two points along
the $x$-axis, $\phi (0)=\phi (a)=0$, then the Greens function between the
plates$\footnote{%
The free Greens function for the other ordering is simply obtained from the
symmetry $G(x,x^{^{\prime }})=G(x^{^{\prime }},x).$}$ for $a>x>x^{^{\prime
}}>0$ is immediately seen to be $\sin (kx^{^{\prime }})\sin [k(a-x)]\csc
(ka)/k.$ For the region $\infty >x^{^{\prime }}>x>a$ one wants outgoing
waves as the boundary condition for Eq.\ref{ode} and so the appropriate
Greens function is $\exp [ik(x^{^{\prime }}-a)]\sin [k(x-a)]/k$. The
positive exponential guarantees convergence once a rotation to the imaginary
axis is made, $k\rightarrow iK$, and Eq.\ref{force} immediately yields the
well-known result for the (attractive) force on the top plate$\footnote{%
For convenience, we shall frequently refer to the Dirichlet points as
``conducting plates'' or ``plates''. The reader may wish to consult Ref. 2
for details leading to the result quoted here.}$, 
\begin{equation}
T^{xx}=-\frac{\pi }{24a^{2}}.
\end{equation}

Having reviewed the necessary formalism in a familiar context, let us now
make a non-trivial choice for $V(x).$ By way of mocking up a constant force
directed towards a fixed centre at $x=0$, choose $V(x)=b\left| x\right| $
with $b>0$ and $-\infty <x<\infty $. Intuitively speaking, as a scalar
photon rises it loses energy and undergoes a redshift. The Dirichlet
condition $\phi (a)=0$ will be said to represent a single ``conducting
plate'' placed above the origin at a height $a$. For a translationally
invariant potential the forces on both sides of the plate would cancel. But,
with a position dependent potential, this would not be true. One can try to
use perturbation theory in the ``coupling constant'' $b$ for computing the
net force on the plate. Although this ultimately fails (for reasons to be
discussed soon), it is nevertheless instructive to make an attempt.

At leading order in $V$, the solution to Eq.\ref{ode} is, 
\begin{equation}
G(x,x^{^{\prime }},k)=G_{0}(x,x^{^{\prime }},k)-\int
dyG_{0}(x,y,k)V(y)G_{0}(y,x^{^{\prime }},k),  \label{pert}
\end{equation}
where, $G_{0}(x,x^{^{\prime }},k)$ is the Greens function for $V=0$ and the
appropriate range of arguments, together with boundary conditions
corresponding to outgoing waves. A calculation for real $k$, followed by
rotation to the imaginary $K$ axis, yields the force just below and just
above the plate at $x=a$, 
\begin{eqnarray}
T_{below}^{xx} &=&\int_{0}^{\infty }\frac{dK}{2\pi }\left[ -K+b(\frac{%
1-2Ka-2e^{-2Ka}}{4K^{2}})\right] , \\
T_{above}^{xx} &=&\int_{0}^{\infty }\frac{dK}{2\pi }\left[ -K-b(\frac{1+2Ka}{%
4K^{2}})\right] .
\end{eqnarray}
The net force is, 
\begin{equation}
T^{xx}=T_{below}^{xx}-T_{above}^{xx}=b\int_{0}^{\infty }\frac{dK}{2\pi }%
\frac{1-e^{-2Ka}}{2K^{2}}.  \label{net}
\end{equation}
Although the linearly divergent integrals have cancelled, there is clearly
an infrared divergence present as $K\rightarrow 0$. It is not hard to
understand its origin: in arriving at Eq.\ref{pert} we have implicitly
assumed that $k^{2}>-\left| V(x)\right| $. Else, oscillatory solutions
cannot exist. But, for a fixed $k$ this condition is violated when $x$
becomes sufficiently large and the unperturbed solution is wholly
unsuitable. To make some sense of Eq.\ref{net} one may think of cutting off
the integral at the lower end with a value $K^{2}\thicksim \left| a\right| b$
in which case $T^{xx}\thicksim \sqrt{\left| a\right| }b^{3/2}$. Of course,
one cannot take this result seriously since the use of perturbation theory
is questionable, as is the imposition of an arbitrary infrared cutoff.
Nevertheless, it is interesting to see that the force thus estimated is
positive, increases with the distance of the plate away from the origin, and
is non-analytic in the strength of the external potential.

It is essential to solve the problem exactly. Fortunately, for the simple
potential we have chosen this is possible. Only the Green's function near
the plate at $x=a$ (with $a>0$) is needed. To proceed, first consider the
region for $0<a<x^{^{\prime }}<x.$ Define a Euclidean dimensionless momentum 
$\kappa $, $k=ib^{1/3}\kappa $. Eq.\ref{ode} becomes, 
\begin{equation}
\left[ \frac{d^{2}}{dy^{2}}+y\right] G(y,y^{^{\prime }},\kappa
)=-b^{-1/3}\delta (y-y^{^{\prime }}),  \label{odey}
\end{equation}
\FRAME{ftbpF}{4.7954in}{3.2949in}{0pt}{}{}{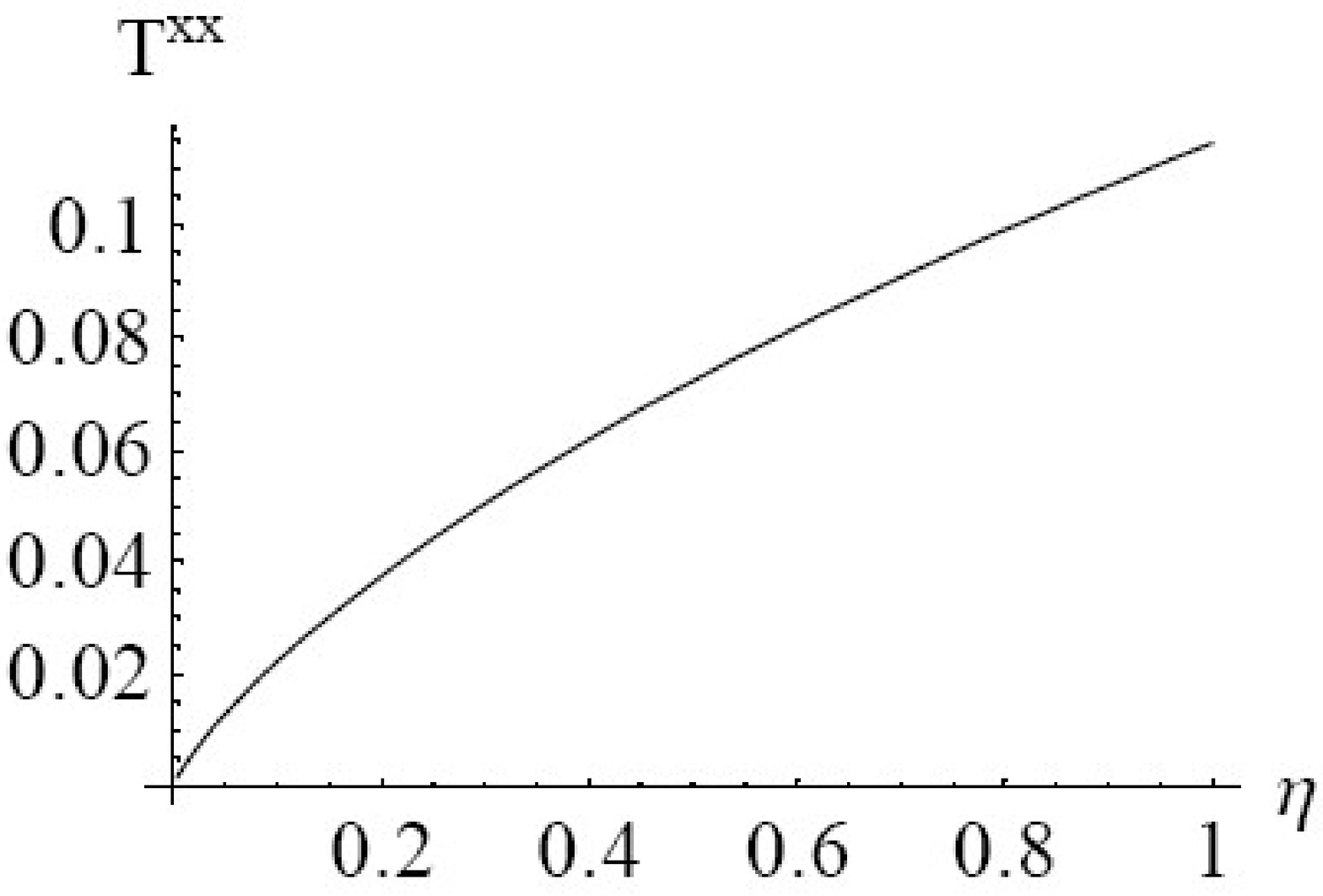}{\special{language
"Scientific Word";type "GRAPHIC";maintain-aspect-ratio TRUE;display
"USEDEF";valid_file "F";width 4.7954in;height 3.2949in;depth
0pt;original-width 344.8125pt;original-height 236.375pt;cropleft "0";croptop
"1";cropright "1";cropbottom "0";filename 'Casfig.ps';file-properties
"XNPEU";}}$y=\kappa ^{2}+(x/a)\eta ^{\frac{1}{3}}$ where $\eta =ba^{3}.$
Both $y$ and $\eta $ are positive and dimensionless. The solutions of $%
G^{\prime \prime }+yG=0$ are the Airy functions, $Ai(y)$ and $Bi(y)$ and the
outgoing wave condition requires that $Bi(y)$ be excluded for $x>a$. The
Green's function in this region is, 
\begin{equation}
\pi a\eta ^{-1/3}Ai(\kappa ^{2}+\frac{x}{a}\eta ^{\frac{1}{3}})\frac{%
Ai(\kappa ^{2}+\eta ^{\frac{1}{3}})Bi(\kappa ^{2}+\frac{x^{^{\prime }}}{a}%
\eta ^{\frac{1}{3}})-Ai(\kappa ^{2}+\frac{x^{^{\prime }}}{a}\eta ^{\frac{1}{3%
}})Bi(\kappa ^{2}+\eta ^{\frac{1}{3}})}{Ai(\kappa ^{2}+\eta ^{\frac{1}{3}})}.
\end{equation}
From this, and Eq.\ref{force}, $T_{above}^{xx}$ follows,

\begin{eqnarray}
T_{above}^{xx} &=&\frac{\eta ^{2/3}}{a^{2}}\int_{0}^{\infty }\frac{d\kappa }{%
2\pi }\frac{Ai^{^{\prime }}(\kappa ^{2}+\eta ^{\frac{1}{3}})}{Ai(\kappa
^{2}+\eta ^{\frac{1}{3}})} \\
&=&\frac{\eta ^{2/3}}{a^{2}}\int_{0}^{\infty }\frac{d\kappa }{2\pi }\left[
-\kappa -\frac{\eta ^{\frac{1}{3}}}{2\kappa }-\frac{1}{4\kappa ^{2}}-\cdot
\cdot \cdot \right] .
\end{eqnarray}

In calculating the force on the other side of the force, one needs to
recognize that the arguments of the Airy functions change into $y=\kappa
^{2}-(x/a)\eta ^{\frac{1}{3}}$ for negative $x$. Again, the outgoing wave
condition requires that $Bi(y)$ be excluded for $x<0$. Finally, one requires
continuity of the solution and derivative at $x=0$, as well as the jump
condition imposed by the delta function. This yields the Green's function,
from which the force below the plate is calculated to be: 
\begin{eqnarray}
\!T_{below}^{xx} &=&\frac{\eta ^{2/3}}{a^{2}}\int_{0}^{\infty }\frac{d\kappa 
}{2\pi }\times  \nonumber \\
&&\frac{2Ai(\kappa ^{2})Ai^{^{\prime }}(\kappa ^{2})Bi^{^{\prime }}(\kappa
^{2}+\eta ^{\frac{1}{3}})-Ai^{^{\prime }}(\kappa ^{2}+\eta ^{\frac{1}{3}%
})(Ai^{\prime }(\kappa ^{2})Bi(\kappa ^{2})+Ai(\kappa ^{2})Bi^{^{\prime
}}(\kappa ^{2}))}{Ai(\kappa ^{2}+\eta ^{\frac{1}{3}})(Ai^{\prime }(\kappa
^{2})Bi(\kappa ^{2})+Ai(\kappa ^{2})Bi^{^{\prime }}(\kappa ^{2}))-2Ai(\kappa
^{2})Ai^{^{\prime }}(\kappa ^{2})Bi(\kappa ^{2}+\eta ^{\frac{1}{3}})} 
\nonumber \\
&=&\frac{\eta ^{2/3}}{a^{2}}\int_{0}^{\infty }\frac{d\kappa }{2\pi }\left[
-\kappa -\frac{\eta ^{\frac{1}{3}}}{2\kappa }+\frac{1}{4\kappa ^{2}}-\cdot
\cdot \cdot \right]
\end{eqnarray}
\linebreak Although $T_{above}^{xx}$ and $T_{below}^{xx}$ are separately
divergent at the upper limit (as might be expected from the infinite
pressure of photons striking each surface), $%
T^{xx}=T_{below}^{xx}-T_{above}^{xx}$ is finite. The integrals must be done
numerically. Reinstating $\hbar $ and $c$, $T^{xx}$ can be expressed as, 
\begin{equation}
T^{xx}=\frac{\hbar c}{a^{2}}f(\eta ).
\end{equation}
In Fig.1 we plot $f(\eta )$. The cusp-like behaviour at $\eta =0$ reflects
the non-analyticity and non-perturbative character of the solution. The
force vanishes at $a=0$, monotonically increases with $a$, and is repulsive.

In summary, it has been shown here that one can expect even a single
conducting plate placed in the vacuum to experience a net quantum force. The
force has the same origin as the Casimir effect, i.e is a manifestation of
the zero-point fluctuations of a quantum field. The difference in the
density of normal modes above and below the plate, induced by the
position-dependent external potential, is the responsible mechanism. The
present investigation was performed with a simple, real, scalar field but
one expects a similar effect for the electromagnetic field (or any other
field) as well.

\medskip

\begin{description}
\item  
\begin{center}
${\bf Acknowledgments}$
\end{center}
\end{description}

The author thanks Bob Jaffe for suggesting this problem, for many
discussions and probing questions. He would also like to acknowledge
discussions with John Negele and Antonello Scardicchio, and thanks the
Center for Theoretical Physics at MIT for hospitality during a visit in
summer 2004.


\begin{thebibliography}{9}
\bibitem{Casimir}  H. B. G. Casimir, Kon. Ned. Akad. Wetensch. Proc. 51
(1948) 793.

\bibitem{Milton}  For a recent review, and lucid introduction to Casimir
problems in the wider context, see ``The Casimir Effect: Recent
Controversies and Progress'' by K. A. Milton, J.Phys.A37:R209,2004,
hep-th/0406024.

\bibitem{Elizalde}  E.Elizalde and A.Romeo, ``One Dimensional Casimir Effect
Perturbed by an External Field'', J.Phys A: Math. 30 (1997), 5393.

\bibitem{Jaffe}  See, for example, ``Casimir Energies in Light of Quantum
Field Theory'', by N. Graham, R.L. Jaffe, V. Khemani, M. Quandt, M.
Scandurra, H. Weigel, Phys.Lett.B572:196-201,2003, hep-th/0207205.
\end{thebibliography}
\end{document}